\documentclass[10pt]{article}
\usepackage{epsf} 
\usepackage{amsmath}
\usepackage{amssymb}
\usepackage{epsfig}
\usepackage{latexsym}
\usepackage{amsfonts}
\usepackage{graphicx}%
\usepackage{varioref}
\usepackage{ifthen}
\begin{document}
\parindent 0mm 
\setlength{\parskip}{\baselineskip} 
\pagenumbering{arabic} 
\setcounter{page}{1}
\mbox{ }
\rightline{UCT-TP-295/2013}
\newline
\newline
\rightline{May 2013}
\newline
\begin{center}
\Large \textbf{INTRODUCTION TO QCD SUM RULES}
\end{center}
\begin{center}
C. A. DOMINGUEZ
\end{center}
\begin{center}
Centre for Theoretical and Mathematical Physics, and Department of Physics,  
University of Cape Town, Rondebosch 7700, South Africa
\end{center}
\begin{center}
\textbf{Abstract}
\end{center}
A general, and very basic introduction to QCD sum rules is presented, with emphasis on recent issues to be described at length in other papers in this volume of Modern Physics Letters A. Collectively, these papers constitute the proceedings of the {\it{International Workshop on Determination of the Fundamental Parameters of QCD}}, Singapore, March 2013.
\noindent
\section{Introduction}
Quark and gluon confinement in Quantum Chromodynamics (QCD) precludes direct experimental measurements of the fundamental QCD parameters, i.e. the strong interaction coupling and the quark masses. Hence, in order to determine these parameters analytically one needs to relate them to experimentally measurable quantities. Alternatively,  simulations of QCD on a lattice provide increasingly accurate numerical values for these parameters, but little if any insight into their origin. The latter may be obtained from an analytical approach which relies on the intimate relation between QCD Green functions, in particular their Operator Product Expansion (OPE) beyond perturbation theory, and their hadronic counterparts. This relation follows from Cauchy's theorem in the complex (squared) energy plane (quark-hadron duality), and is  collectively known as the QCD sum rule technique \cite{QCDSR}. In addition to producing numerical values for the QCD parameters, this method provides a detailed breakdown of the relative impact of the various dynamical contributions. For instance, the strong coupling at the scale of the $\tau$-lepton mass  essentially follows from the relation between the experimentally measured $\tau$ ratio, $R_\tau$, and a contour integral involving the perturbative QCD (PQCD) expression of the $V+A$ correlator. This is the cleanest, most transparent, and model independent determination of the strong coupling \cite{PDG}-\cite{alpha4}. It also allows to gauge the impact of each individual term in PQCD, up to the currently known five-loop order.
Similarly, in the case of the quark masses one considers a QCD correlation function which on the one hand involves the quark masses and other QCD parameters, and on the other hand it involves a hadronic spectral function, experimentally measurable in some cases. Using Cauchy's theorem to relate both representations, the quark masses  become a function of QCD parameters, e.g. the strong coupling, some vacuum condensates reflecting confinement, etc., and  hadronic parameters. The virtue of this approach is that it provides a breakdown of each contribution to the final value of the quark masses. More importantly, it allows to tune the relative weight of each of these contributions by introducing suitable integration kernels. This feature is very important to either quench or enhance contributions which are either poorly or well known, respectively.\\ 
\section{Operator product expansion beyond perturbation theory}
The OPE beyond perturbation theory in QCD, one of the two pillars of the sum rule technique, is an effective
tool to introduce quark-gluon confinement dynamics. It is not a model, but rather a parametrization of quark and gluon propagator corrections due to confinement, done in a rigorous renormalizable quantum field theory framework. Let us consider a typical object in QCD in the form of the two-point function, or current correlator
\begin{equation}
\Pi(q^2)\,=\,i\; \int \,d^4 x \; e^{iqx} \; <0|\,T(J(x)\,J(0))\,|0 >,
\end{equation}
where the local current $J(x)$ is built from the quark and gluon fields entering the QCD Lagrangian. Equivalently, this current can also be written in terms of hadronic fields with identical quantum numbers. A relation between the two representations follows from Cauchy's theorem in the complex energy (squared) plane. This is often referred to as quark-hadron duality, the second pillar of the QCD sum rules (QCDSR) method to be discussed in the next section.
The QCD correlator, Eq.(1),  contains a perturbative piece (PQCD), and a non perturbative one mostly reflecting quark-gluon confinement.  Since confinement has not been proven analytically and exactly in QCD, its effects  can only be introduced effectively, e.g. by parameterizing quark and gluon propagator corrections in terms of vacuum condensates. This is done as follows. In the case of the quark propagator
\begin{equation}
S_F (p) = \frac{i}{\not{p} - m}\;\;\Longrightarrow \;\;\frac{i}{\not{p} - m + \Sigma(p^2)} \;, 
\end{equation}
the  propagator correction $\Sigma(p^2)$  contains the information on confinement, a purely non-perturbative effect. One expects this correction to peak at and near the quark mass-shell, e.g. for $p \simeq 0$ in the case of light quarks. This effect is then parameterized in terms of the quark condensate $\langle 0| \bar{q}(0) q(0) | 0 \rangle$.
Similarly, in the case of the gluon propagator 
\begin{equation}
D_F (k) = \frac{i}{k^2}\;\;\Longrightarrow \;\;\frac{i}{k^2 + \Lambda(k^2)} \;,
\end{equation}
the propagator correction $\Lambda(k^2)$ will peak at $k\simeq 0$, and the effect of confinement in this case can be parameterized by the gluon condensate $\langle 0| \alpha_s\; \vec{G}^{\mu\nu} \,\cdot\, \vec{G}_{\mu\nu}|0\rangle$.
In addition to the quark and the gluon condensate there is a plethora of higher order condensates entering the OPE of the current correlator at short distances, i.e.
\begin{equation}
\Pi(q^2)|_{QCD}\,=\, C_0\,\hat{I} \,+\,\sum_{N=0}\;C_{2N+2}(q^2,\mu^2)\;\langle0|\hat{O}_{2N+2}(\mu^2)|0\rangle \;,
\end{equation}
where $\mu^2$ is the renormalization scale, and where the Wilson coefficients in this expansion, 
$ C_{2N+2}(q^2,\mu^2)$,  depend on the Lorentz indexes and quantum numbers of $J(x)$ and  of the local gauge invariant operators $\hat{O}_N$ built from the quark and gluon fields. These operators are ordered by increasing dimensionality and the Wilson coefficients, calculable in PQCD, fall off by corresponding powers of $-q^2$. In other words, this OPE achieves a factorization of short distance effects encapsulated in the Wilson coefficients, and long distance dynamics present in the vacuum condensates.
Since there are no gauge invariant operators of dimension $d=2$ involving the quark and gluon fields in QCD, it is normally assumed that the OPE starts at dimension $d=4$. This is supported by results from QCD sum rule analyses of $\tau$-lepton decay data, which show no evidence of $d=2$ operators \cite{C2a}-\cite{C2b}.
The unit operator $\hat{I}$ in Eq.(4) has dimension $d=0$ and $C_0 \hat{I}$ stands for the purely perturbative contribution. The Wilson coefficients as well as the vacuum condensates depend on the renormalization scale. For light quarks, and for the leading $d=4$ terms in Eq.(4), the $\mu^2$ dependence of the quark mass cancels the corresponding dependence of the quark condensate, so that this contribution is a renormalization group (RG) invariant. Similarly, the gluon condensate is also a RG invariant, hence once determined in some channel these condensates can be used throughout. For light quarks these statements are correct up to quartic quark-mass terms, as there are some issues with the cancellation of logarithmic quark-mass singularities \cite{m41}-\cite{m42}.
The numerical values of the vacuum condensates cannot be calculated analytically from first principles as this would be tantamount to solving QCD exactly.
One exception is that of the quark condensate which enters in the Gell-Mann-Oakes-Renner relation, a QCD low energy theorem following from the global chiral symmetry of the QCD Lagrangian \cite{GMOR}. Otherwise, it is possible to extract values for the leading vacuum condensates using QCDSR together with experimental data, e.g. $e^+ e^-$ annihilation into hadrons, and hadronic decays of the $\tau$-lepton. Alternatively, as lattice QCD  improves in accuracy it should become a valuable source of information on these condensates.\\
\section{Quark-hadron duality and finite energy QCD sum rules}
Turning to the hadronic sector, bound states and resonances appear in the complex energy (squared) plane (s-plane) as poles on the real axis, and singularities in the second Riemann sheet, respectively. All these singularities lead to a discontinuity across the positive real  axis. Choosing an integration contour as shown in Fig. 1, and given that there are no other singularities in the complex s-plane, Cauchy's theorem leads to the finite energy sum rule (FESR)
\begin{figure}[ht]
\begin{center}
  \includegraphics[height=.18\textheight]{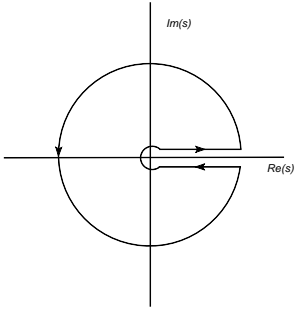}
  \caption{\footnotesize{Integration contour in the complex s-plane. The discontinuity across the real axis brings in the hadronic spectral function, while integration around the circle involves the QCD correlator. The radius of the circle is $|s_0|$, the onset of QCD.}}
\label{fig:figure1}
\end{center}
\end{figure}
\begin{equation}
\int_{\mathrm{sth}}^{s_0} ds\; \frac{1}{\pi}\; p(s) \;Im \,\Pi(s)|_{HAD} \; = \; -\, \frac{1}{2 \pi i} \; \oint_{C(|s_0|) }\, ds \;p(s) \;\Pi(s)|_{QCD} \;,
\end{equation}
where the kernel $p(s)$ is an arbitrary (analytic) function, $s_{th}$ is the hadronic threshold, and the finite radius of the circle, $|s_0|$, is large enough for QCD and the OPE to be used on the circle. 
Physical observables determined from FESR should be independent of $s_0$. In practice, though, 
this  is not exact, and there is usually a region of stability where
observables are fairly independent of $s_0$, typically somewhere inside the range $s_0 \simeq 1 - 4 \; \mbox{GeV}^2$. 
Equation (5) is the mathematical statement of what is usually referred to as quark-hadron duality. Since PQCD is not valid in the time-like (resonance) region ($s \geq 0$), in principle there is a possibility of problems on the circle near the real axis (duality violations), to be discussed shortly (this issue was identified very early in \cite{Shankar} long before the present formulation of QCDSR).
The right hand side  of this FESR involves the QCD correlator which is expressed in terms of the OPE as in Eq.(4). The left hand side involves the hadronic spectral function, which may contain a 
ground state pole,  followed by  resonances which merge smoothly into the hadronic continuum above some threshold $s_0$. This continuum is expected to be well represented by PQCD if $s_0$ is large enough.\\ 
Next, let us consider  an application where the integration kernel $p(s)$ in Eq.(5) is of great importance. For the difference between the vector and axial-vector correlators (V-A)  the FESR, Eq.(5), with $p(s)=1$ and $N=0$ becomes the (finite energy) first Weinberg sum rule (WSR), which can be confronted with data from $\tau$-decay \cite{ALEPH},
\begin{equation}
W_1(s_0) \equiv f_\pi^2 = \int_0^{s_0} \, ds\; p(s)\;{\mbox{Im}}\; [\Pi_V(s) - \Pi_A(s)] \;,
\end{equation}
where $\Pi_{V,A}(s)$ are the vector and the axial-vector correlators, respectively, and $p(s)=1$ in the original WSR.
\begin{figure}[ht]
\begin{center}
  \includegraphics[height=.20\textheight]{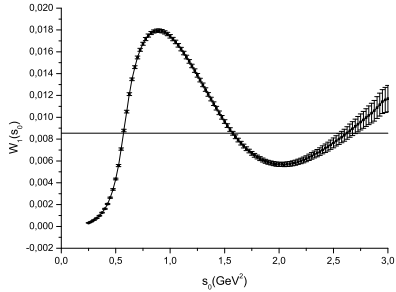}
  \caption{ \footnotesize{Results for $f_\pi^2$  (in units of $ {\mbox{GeV}}^2$) from the first Weinberg sum rule treated as a FESR.  The straight line  is the experimental value of $f_\pi^2$, and the points are the integrated ALEPH data for the V-A correlator \cite{ALEPH}.}}
  \label{fig:figure2}
  \end{center}
\end{figure}
\begin{figure}[ht]
\begin{center}
  \includegraphics[height=.20\textheight]{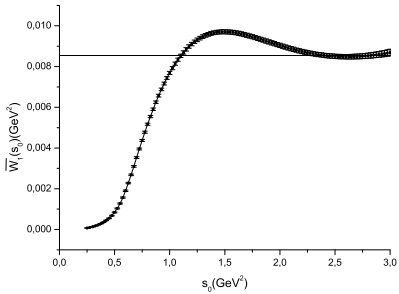}
  \caption{ \footnotesize{Results for $f_\pi^2$ from the first Weinberg sum rule treated as a FESR with a pinched integration kernel $p(s)=(1-s/s_0)$. The straight line  is the experimental value of $f_\pi^2$, and the points are the integrated ALEPH data for the V-A correlator \cite{ALEPH}.}}
  \label{fig:figure3}
  \end{center}
\end{figure}
As seen from Fig. 2 the agreement is rather poor, except possibly near the end point. At first sight, this may be interpreted as a signal for quark-hadron duality violations near the real axis, even at this high enough energy. In fact, it has been known for quite some time that  the Weinberg (chiral) sum rules are not saturated by the $\tau$ decay data unless one introduces {\it pinched} integration kernels, e.g.  $p(s) = [1 - (s/s_0)]^{(N+1)}$ \cite{PINCH1}-\cite{PINCH2}. In Fig.3 we show the dramatic improvement after introducing the lowest order ($N=0$) pinched kernel. Unfortunately, the $\tau$-lepton is not massive enough to probe higher energy regions. In spite of this it is still possible to explore a wider  energy range by introducing as integration kernel a polynomial $p(s) \equiv P(s, s_0, s_1)$ tuned to eliminate the (unknown) hadronic contribution to the integral between $s_1$ and $s_0 \geq s_1$, where $s_1$ is at or near the end point of the data. It has been shown \cite{FPI} that in the axial-vector channel the optimal degree of $P(s)$ is the simplest, i.e. the linear function
\begin{equation}
P(s,s_0,s_1)=1-\frac{2s}{s_{0}+s_{1}} \,,
\end{equation}
so that
\begin{equation}
 \mbox{constant} \times \int_{s_1}^{s_0} P(s,s_0,s_1) ds  = 0\,.
\end{equation}
In this case the complete FESR (in the axial-vector channel) becomes a linear combination of a dimension-two and a dimension-four FESR, i.e.
\begin{eqnarray}
2 \, f_\pi^2 &=& - \int_{0}^{s_{1}} ds \, P(s) \, \frac{1}{\pi}\, Im \,\Pi(s)_A|_{DATA}
+ \frac{s_0}{4 \pi^2} \left[ M_2(s_0) - \frac{2 s_0}{s_0+s_1} M_4 (s_0) \right] \nonumber \\[.3cm]
&+& \frac{1}{4 \pi^2} \left[ C_2 \langle \hat{O}_2 \rangle +\frac{2}{s_0+s_1} C_4 \langle \hat{O}_4 \rangle \right] \, + \Delta(s_0)\,,
\end{eqnarray}
where the pion pole has been separated from the data, the chiral limit is understood, and 
the dimensionless PQCD moments $M_{2N+2}(s_0)$ are given by
\begin{equation}
M_{2N+2}(s_0) = \frac{1}{s_0^{(N+1)}} \, \int_0^{s_0}\, ds\,s^N \, \frac{1}{\pi} \, Im \, \Pi(s)|_{PQCD}\;.
\end{equation}
The term $\Delta(s_0)$ is the error being made by assuming that the data is constant in the interval $s_1 - s_0$. It is possible to estimate this error which turns out to be two to three orders of magnitude smaller than  $2 f_\pi^2$ entering the left hand side of Eq.(9) \cite{FPI}. 
As can be seen from Fig. 4 the FESR Eq.(9) shows an excellent consistency between QCD and the $\tau$ data in the axial-vector channel in a remarkably wide region $s_0 \simeq 4 \, -\,10\, \mbox{GeV}^2$. A similar consistency is also found in the vector channel, using the same integration kernel, and where QCD is now confronted with zero (there is no pole in this channel). These results show either no evidence for quark-hadron duality violations in these channels, and at these energies, or, if they are present, they indicate a suppression due to the integration kernel  (some model dependent analyses claim the existence of duality violations \cite{PICH3a}-\cite{PICH3c}).\\ 
\begin{figure}[ht]
\begin{center}
  \includegraphics[height=.25\textheight]{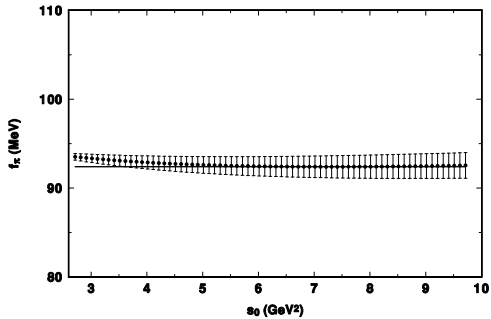}
  \caption{\footnotesize{Results for $f_\pi$ from the FESR in the axial-vector channel, Eq.(9).}}
  \label{fig:figure4}
  \end{center}
\end{figure}
\section{Unveiling systematic uncertainties}
Systematic uncertainties in QCDSR, understood as {\it{a-priori}} fully unknown errors, arise from two sources, the hadronic and  the QCD sector. In all cases where they were successfully unveiled, these errors acted in only one direction. The difficult task of unveiling these uncertainties has taken many years to accomplish, with most of the success having taken place in recent times. Beginning with the hadronic sector, historically the first applications of QCDSR involved hadronic resonances in the zero-width approximation. This was followed by finite width parameterizations, albeit with no model-independent threshold constraints. The use of chiral perturbation theory to constrain resonance threshold behaviour was first proposed in \cite{CAD_CPT}, and enforced  in the $3 \pi$ channel, and later in the $K \pi \pi$ channel \cite{KPIPI}-\cite{SB1}. In the case of the light-quark pseudoscalar correlator, involved in the determination of the three light-quark masses, the above constraint takes care of only a small part of the systematic uncertainty. This arises from the lack of full experimental information on the hadronic resonance spectral function, i.e. only the pseudoscalar meson poles are fully known from experiment. The presence of at least two radial excitations of the pion and the kaon has been established, and their masses and widths are known. However, this is hardly enough to reconstruct the full hadronic spectral function. In fact, non-resonant background, and interference are impossible to guess. A major step in reducing considerably this systematic uncertainty was made in \cite{DNRS}-\cite{DNRS2}, with the introduction of an integration kernel which vanishes at the peak of each of the two resonances, i.e.
\begin{equation}
p(s) = (s - M_1^2) (s - M_2^2) \;,
\end{equation}
where $M_{1,2}$ are the masses of the first two pseudoscalar resonances.
\begin{figure}
[ht]
\begin{center}
  \includegraphics[height=.25\textheight]{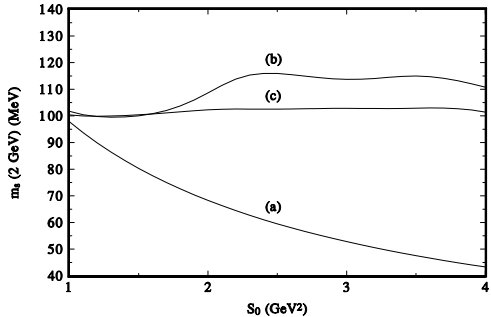}
\caption{\footnotesize{The result for $\bar{m}_s(2 \, {\mbox{GeV}})$ from \cite{DNRS} using only the kaon pole, curve (a), the full hadronic spectral function with no integration kernel ($p(s)=1$), curve (b), and with $p(s)$ as in Eq.(11), curve (c).}}
\end{center}
\end{figure}
Another systematic uncertainty is due to the value of the threshold for PQCD, $s_0$.  For instance, in the case of light-quark mass determinations, since the quark mass must be independent of the value of the Cauchy radius, $|s_0|$, (provided it is large enough for QCD to be valid), it has been traditional to seek  as much stability as possible in this mass against changes in $|s_0|$. Figure 5 shows results for $\bar{m}_s(2 \, {\mbox{GeV}})$ obtained in \cite{DNRS} using only the kaon pole, curve (a), and the full hadronic spectral function (kaon pole plus two radial excitations), but with no integration kernel, i.e. $p(s) = 1$, curve (b). Since this result is reasonably stable for $s_0 \simeq 2.2 - 4.0 \; {\mbox{GeV}}^2$, one would have concluded that $\bar{m}_s(2 \, {\mbox{GeV}}) \simeq 110 - 118 \; {\mbox{MeV}}$, with an additional error due to other sources. However, using the integration kernel $p(s)$, Eq.(11), thus quenching the contribution of the resonance sector, leads to the considerably lower result $\bar{m}_s \simeq 100\; {\mbox{MeV}}$, thus unveiling a $10 - 20 \%$ systematic uncertainty from the hadronic sector, and acting in only one direction.\\
Turning to the QCD sector, a potential source of serious systematic uncertainty stems from correlators with poor PQCD convergence, e.g. the light-quark pseudoscalar correlator used to determine the strange-quark mass
\begin{equation}
\psi_{5} (q^{2})   = i \, \int\; d^{4}  x \; e^{i q x} \; 
<|T(\partial^\mu A_{\mu}(x) \;, \; \partial^\nu A_{\nu}^{\dagger}(0))|> \;,
\end{equation}
where $\partial^\mu A_{\mu}(x) = (m_s + m_{ud}) :\overline{s}(x) \,i \, \gamma_{5}\, u(x):\;$ is the divergence of the  axial-vector current, and $m_{ud} \equiv (m_u + m_d)/2$. The second derivative of $\psi_5(q^2)$ at a scale $\mu^2 = Q^2 \equiv -q^2$  in the $\overline{MS}$ scheme, to five-loop order in PQCD \cite{CHET_K} is given by
\begin{equation}
\psi_{5}^{''} (Q^{2})^{PQCD} = \frac{3}{8 \, \pi^2} \frac{(\overline{m}_s + \overline{m}_{ud})^2}{Q^2}
\left[1 +  3.7\, a + 14.2\, a^2 + 77.4\, a^3 + 512.0\, a^4\right] \;,
\end{equation}
where $a \equiv \alpha_s(Q^2)/\pi$. This behaviour is already providing a strong hint of a potential systematic uncertainty. If present, it could change both the central value as well as the error in 
the result for the strange quark mass, to wit. Let us define
\begin{equation}
\delta_5^{QCD}(s_0) = - \frac{1}{2\pi i} \oint_{C(|s_0|)} ds \; p_i(s) \;\psi_{5}^{OPE}(s) \;,
\end{equation}
where for convenience  the quark masses have been factored out of $\psi_5^{QCD}(s)$, and thus $\psi_{5}^{OPE}(s)$ is  the remainder in PQCD plus power corrections in the OPE. The perturbative QCD expansion of $\delta_5^{PQCD}$, Eq.(14), with the integration kernel $p_1(s)$, Eq. (11), and for $s_0 = 4.2\, {\mbox{GeV}}^{2}$ (with $\mu=\sqrt{s_0}$) is given by
\begin{equation}
\delta_5^{PQCD}= 0.23 \, {\mbox{GeV}}^{8} \left[1 + 2.2 \,\alpha_{s} + 6.7\,\alpha_{s}^2 + 19.5 \,\alpha_{s}^3 + 56.5\, \alpha_{s}^4\right]\;,
\end{equation} 
which after replacing a typical value of $\alpha_s$ leads to all terms beyond the leading order to be roughly the same, e.g. for $\alpha_s = 0.3$ the result is
\begin{equation}
\delta_5^{PQCD}= 0.23 \, {\mbox{GeV}}^{8} \left[1 + 0.65 + 0.60 + 0.53 + 0.46\right]\;,
\end{equation}
which is hardly (if at all) convergent. In fact, judging from the first five terms, this expansion
is worse behaved than the non-convergent harmonic series. Once the systematic uncertainty is unveiled, it is possible to to find ways of reducing its impact, e.g. by accelerating the PQCD convergence \cite{SB1},\cite{KS}. In the case of the strange-quark mass determination, its value turns out to be some 20\% smaller after taking care of this systematic uncertainty \cite{SB1}.\\
Heavy-quark masses (charm and bottom) are, in principle, not affected by the lack of information on the hadronic spectral function, as there is experimental data from $e^+ e^-$ annihilation into hadrons. However, in some regions there are conflicting results from different experiments,  and in other regions the errors are too large, or there are simply no data at all \cite{DENNIG}-\cite{ACS}. The introduction of integration kernels that quench or enhance regions of poor or precise data, respectively, has allowed for a considerable reduction in the overall uncertainty in the heavy-quark masses, see e.g. \cite{SB2}-\cite{JK}. In fact, QCDSR results are now competing in accuracy with lattice QCD determinations \cite{LQCD}.\\
Another example of the successful use of integration kernels is the theoretical determination of the muon anomaly, $g-2$. Indeed, after quenching the $e^+ e^-$ data in conflictive regions, the value of $g-2$ becomes closer  to experiment, thus making the case for effects beyond the Standard  Model less compelling \cite{SB3}.
\section{Acknowledgments}
The author wishes to thank Professor K. K. Phua, and the Institute of Theoretical Physics, Nanyang Technological University, for their generous support of this workshop. This work was supported in part by NRF (South Africa) and Alexander von Humboldt Foundation (Germany).

\end{document}